\documentclass[trackchanges]{aastex701}

\newcommand{\gray}{$\gamma$-ray}

\begin{document}

\title{On the Sensitivity of Fermi-LAT Maximum Likelihood Stacking}

\author[orcid=0000-0002-0472-9195,sname='O. K. Henry']{Owen K. Henry}
\affiliation{The Graduate Center, Physics Program, City University of New York, 365 Fifth Ave., New York, NY 10016, USA}
\affiliation{Department of Astrophysics, American Museum of Natural History, Central Park West at 79th Street, New York, NY 10024, USA}
\email[show]{ohenry@amnh.org}  

\author[orcid=0000-0001-9139-0945, sname='Paglione']{Timothy A. D. Paglione}
\affiliation{Department of Earth \& Physical Sciences, York
College, City University of New York, 94-20 Guy R. Brewer Blvd., Jamaica, NY 11451, USA}
\affiliation{Department of Astrophysics, American Museum of Natural History, Central Park West at 79th Street, New York, NY 10024, USA}
\email{}

\author[orcid=0000-0003-3835-8115, sname='Tan']{Joshua Tan}
\affiliation{Department of Natural Sciences, LaGuardia Community College, City University of New York, 31-10 Thomson Ave, Long Island City, NY 11101}
\affiliation{Department of Astrophysics, American Museum of Natural History, Central Park West at 79th Street, New York, NY 10024, USA}
\email{}

\author[orcid=0009-0008-4599-2935, sname='Zurek']{David Zurek}
\affiliation{Department of Astrophysics, American Museum of Natural History, Central Park West at 79th Street, New York, NY 10024, USA}
\email{}

\author[orcid=0009-0006-9984-3885, sname='Eysselinck']{Thyra Eysselinck}
\affiliation{Department of Physics \& Astronomy, George Mason University, C4400 University Drive, MSN 3F3, Fairfax, VA 22030, USA}
\email{}

%\keywords{\uat{Galaxies}{573} --- \uat{Cosmology}{343} --- \uat{High Energy astrophysics}{739} --- \uat{Interstellar medium}{847} --- \uat{Stellar astronomy}{1583} --- \uat{Solar physics}{1476}}

\begin{abstract}
    Stacking sub-threshold \gray\ sources has become common practice for understanding high-energy emission processes at the population level. In this procedure, one co-adds log-likelihood ratios, called TS values (TS = 2ln$(L_1/L_0)$), for a source class of interest. It is best practice to compare the population of sub-threshold targets against a population of control field test sources (CFs) selected to represent the unmodeled \gray\ background. Determining the significance of a TS measurement stack has typically been difficult. In this work, we find the necessary cumulative TS value ($\Sigma$TS) of a sub-threshold population of $N$ targets to achieve a $5\sigma$ separation from a like-sized population of CFs. We find that the $\Delta (\Sigma$TS) needed to achieve a $5\sigma$ separation is $\sim 70$. In taking this measurement, we characterize the source distribution corresponding to the measured $\Sigma$TS for a significantly detected stack and even predict new individual source detections in larger stacks. 
\end{abstract}

\section{Introduction}
Stacking sub-threshold \gray\ sources has become common practice for understanding high-energy emission processes at the population level. In this procedure, one co-adds log-likelihood ratios called TS values (TS = 2ln$(L_1/L_0)$) for a source class of interest. The TS value is a Bayesian statistic that compares the model likelihood of a \gray\ point source present in a region of interest (ROI), $L_1$, versus the null hypothesis that there is no source, $L_0$  \citep{1996ApJ...461..396M}. Stacking has been utilized for studies of extreme blazars \citep{2019ApJ...882L...3P}, the extragalactic background light \citep{2020ApJ...894...88A}, FR0 radio galaxies \citep{2024ApJ...971...84K}, flare stars \citep{2024MNRAS.531.3215S}, pulsars \citep{2023MNRAS.524.5854S}, globular clusters \citep{2024MNRAS.535..434H}, among many more \citep{2012A&A...547A.102H,2013A&A...560A..64H, 2021MNRAS.507.4564P, 2023AAS...24112201M, 2023HEAD...2030003K, 2025arXiv251113818D}. %In many stacking studies, it has become standard practice to compare the population of sub-threshold targets against a population of control field test sources (CFs) selected to represent the \gray\ background. 
It is best practice to compare the population of sub-threshold targets against a population of control field test sources (CFs) selected to represent the unmodeled \gray\ background. %Given the procedure for analyzing Fermi-LAT \gray s, where observed photon data are compared to emission models, the unmodeled background will contain sources. 
The CF test source TS distributions deviate from the $\chi^2/2$ theoretical null distribution, indicating a systematic excess of emission beyond statistical variance is present in the Fermi dataset. Because of this, determining the significance of a TS measurement stack has typically been difficult.  %Naively, the $\sqrt{\Delta \text{TS}}$ between the cumulative TS ($\Sigma$TS) of the targets and CFs %and targets 
%has been assumed as a proxy for stack significance ($\sigma$), but this has often been found to be an overestimate. %For instance, the stack separation in \citet{2023MNRAS.524.5854S} yields a $\sim17 \sigma$ from this convention. 
\citet{2024MNRAS.535..434H} presented a novel method to estimate the stack separation's significance between targets and CF test sources utilizing the Kullback-Leibler divergence \citep{10.1214/aoms/1177729694}. The method uses synthesized populations of point sources in a model ROI. With the Fermi Gamma-Ray Space Telescope as the only operational GeV observatory for the foreseeable future, stacking surveys and other strategies using Fermi-LAT's archive could become the predominant way to study the \gray\ sky. In this work, we investigate the necessary cumulative TS value ($\Sigma$TS) of a sub-threshold population of $N$ targets to achieve a $5\sigma$ separation from a like-sized population of CFs. %In this work we investigate the $N$ necessary to yield a $5\sigma$ detection of a sub-threshold population against a comparison population of CFs.

\section{Probing stack sensitivity with synthesized source populations} 
We compare a synthesized population of point source targets to a large library of high-latitude CF test sources observed by the Fermi-LAT. 
%real on-sky CFs. 
We test the stack separation significance for population sizes $N$ = 50 to 200 in increments of 25. %To assess the significance of the targets' cumulative TS stack separation from the stack of CFs, we model the $\gamma$-ray photon counts from a sub-threshold population as in \citet{2024MNRAS.535..434H}. 
The  %
significance between the target and CF stacks is assessed by modeling the difference in their cumulative TS values, $\Delta(\Sigma{\text{TS}})$. %from a %parame ter grid defining the population of model sources. %with fluxes that follow a power-law distribution. 
The model assumes that the photon counts per pixel follow a Poisson distribution, the high-latitude \gray\ background is isotropic (20 counts), and the spatial distribution of point source counts is a 2-D Gaussian with a FWHM of $0.5^{\circ}$. We assume a source population that follows a power law flux distribution proportional to $S^{-\alpha}$, where $S$ is the photon flux, and $\alpha$ is the power law index. The normalization depends on the signal-to-noise ratio, which compares the total counts on source (source plus background) to the total background counts in the same number of pixels. The ratio of source to background counts ($g$) is necessarily well below unity for a sub-threshold source. Our objective is to identify the cumulative TS value of the target stack as a function of population size $N$ that %will 
yields a $5\sigma$ separation from the CF test source stack according to the Kullback-Leibler divergence. We test the target stack's divergence from the CF test source stack by synthesizing a distribution of model point sources parameterized by a $g_{max}$, which we define as the maximum extent of the distribution in the $g$ domain, and a power-law index of $\alpha$. We center this parameter space on the slope of the high-latitude 4FGL source distribution \citep[$\alpha_{4FGL} = -0.873$,][]{2024MNRAS.535..434H}. 

\begin{figure}[h]
    \centering
    \includegraphics[width=0.8\linewidth]{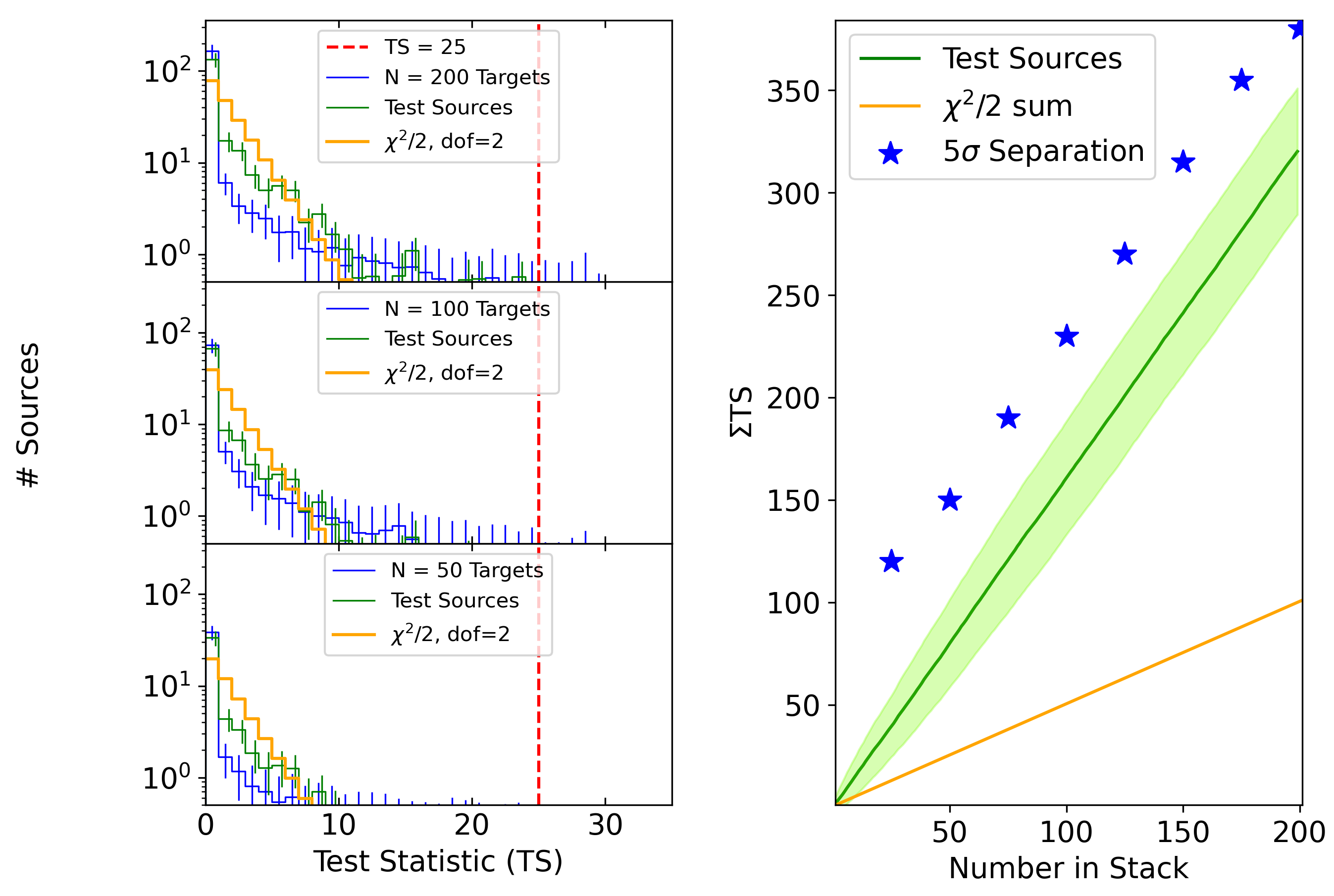}
    \caption{\textit{Left}: A subset of the tested synthesized target distributions (N = 50, 100, 200) plotted with the CF (green) and theoretical null $\chi^2$ distributions (orange). \textit{Right}: The cumulative TS values of the CFs and the theoretical null. % $\chi^2$ distribution. 
    The blue stars represent the cumulative TS for the synthesized target distribution required for a $5\sigma$ separation from the CFs. }
    \label{fig:plot}
\end{figure}

We create an ensemble of source populations over a grid of $\alpha$ and $g_{max}$. %Each grid point defines the distribution shape for 10,000 sources. 
Using $S(\alpha, g_{max})$ as a probability distribution with \texttt{scipy.stats.powerlaw.rvs} \citep{2020SciPy-NMeth} we create 10,000 sources at each grid point.
%Each grid point provides input to a flux distribution function, $S(\alpha, g_{max})$, realized by \texttt{scipy.powerlaw.rvs} \citep{2020SciPy-NMeth}. 
%To generate synthetic data for our toy model, we draw a pixel count from a Poisson distribution. %The model ROI consists of a central source modeled as a 2-D Gaussian counts distribution atop a 20-count background. 
For each pair of $\alpha$ and $g_{max}$, we sample the synthetic source distribution to $N$ and calculate TS values from this sampled population using pixel counts drawn from a Poisson distribution. We sum these TS values to obtain the $\Sigma$TS for the synthetic population. For each grid point, we repeat this process 1,000 times and then average. %We map this value to an array in the $\alpha$-$g_{max}$ space. %10,000 TS values are generated. For each population size, $N$, we sample the respective values and then sum. We do this sampling 10,000 times, and then average, yielding a $\Sigma$TS grid in $\alpha$-$g_{max}$ space. 
%Our population of 800 CFs is sampled in the same way. 
%
%
%We use this method to characterize the stack of our CF test sources. 
With this method, we can estimate the significance of the target stack over the CFs using the KL divergence with \texttt{scipy.special.kl\_div} and the $\sqrt{\Delta (\Sigma{\rm TS})}$. In this work, we seek the  $\Sigma$TS threshold to achieve 5$\sigma$ separation away from the CF test source stack for various synthesized target source population sizes, $N$. %Using this procedure, we predict the sensitivity of our stacking analysis (Fig. \ref{fig:plot}). 
The left panels of Fig. \ref{fig:plot} show the target and CF TS distributions. Note that for $N \gtrsim 100$, we occasionally predict some sources with TS $ > 25$. In the right panel of Fig. \ref{fig:plot}, we show the cumulative TS for the CFs, the $\chi^2/2$ null, and the $5\sigma$ threshold for each tested $N$. Across all synthesized population sizes, we find that the $\Delta (\Sigma$TS) needed to achieve a $5\sigma$ separation is $\sim 70$. %; greater than the conventional $\sqrt{\Delta \text{TS}}>25$  would suggest. 
%For a sample of 200 CF test sources, we estimate that a cumulative TS of $\sim$380 for a like-sized population of targets is required to achieve a 5$\sigma$ separation from the CF test sources (for which $\Sigma$TS $\approx$ 318) (Fig. \ref{fig:plot}).

In this work, we gained insight into the minimum joint likelihood required to reach a $5\sigma$ stack detection given $N$ co-added ROIs. %the characteristics of stacks for subthreshold source populations of various sizes. 
%We expect in any population of \gray\ sources that there are many which lie beneath the detection threshold. 
%If we stack subthreshold targets, we expect to measure greater significance compared to a comparable stack of CF test sources for $N$ target ROIs. 
%In this work, we found the minimum $\Sigma$TS for a $5\sigma$ stack separation given $N$ co-added ROIs. 
In doing so, we characterize the source distribution corresponding to the measured $\Sigma$TS for a significantly detected stack and even predict new individual source detections in larger stacks. Therefore, a large enough population with one or few detected objects is suitable for a gamma-ray joint likelihood stacking study.

\pagebreak
\bibliography{sample701}{}
\bibliographystyle{aasjournalv7}

\end{document}